\documentclass[prl,aps,nofootinbib,notitlepage,twocolumn]{revtex4-1}

\usepackage{amsmath}
\usepackage{amssymb}
\usepackage{graphicx}

\usepackage{txfonts}
\usepackage{dsfont}

\usepackage[colorlinks=true,linkcolor=blue,citecolor=blue,urlcolor=blue]{hyperref}

\providecommand{\norm}[1]{\lVert#1\rVert}

\providecommand{\tr}{{\rm tr}}
\renewcommand{\phi}{\varphi}

\begin{document}
\title{Remote generation of Wigner-negativity through Einstein-Podolsky-Rosen steering}

\author{Mattia Walschaers} 
\email{mattia.walschaers@lkb.upmc.fr}
\affiliation{Laboratoire Kastler Brossel, Sorbonne Universit\'e, CNRS, ENS-PSL Research University, Coll\`ege de France; 4 place Jussieu, F-75252 Paris, France}
\author{Nicolas Treps}
\affiliation{Laboratoire Kastler Brossel, Sorbonne Universit\'e, CNRS, ENS-PSL Research University, Coll\`ege de France; 4 place Jussieu, F-75252 Paris, France}

\date{\today}

\begin{abstract}
Negativity of the Wigner function is seen as a crucial resource for reaching a quantum computational advantage with continuous variable systems. However, these systems, while they allow for the deterministic generation of large entangle states, require an extra element such as photon subtraction to obtain such negativity. Photon subtraction is known to affect modes beyond the one where the photon is subtracted, an effect which is governed by the correlations of the state. In this manuscript, we build upon this effect to remotely prepare states with Wigner-negativity. More specifically, we show that photon subtraction can induce Wigner-negativity in a correlated mode if and only if that correlated mode can perform Einstein-Podolsky-Rosen steering in the mode of subtraction.  
\end{abstract}

\maketitle

The development of a quantum internet \cite{Kimble:2008aa,Wehner:2018aa} is an important goal in the pursuit of quantum technologies. The key idea of such a quantum network, be it for communication or for distributed computation, is to connect a large number of nodes via quantum entanglement \cite{Cirac:1997aa,Morin:2014aa}. A platform that is particularly promising for such applications is continuous-variable quantum optics, where large entangled graph states can be deterministically produced \cite{Su:12,PhysRevLett.112.120505,cai-2017,Larsen:2019aa,Asavanant:2019aa}. Even though this allows us to produce intricate quantum networks, the resulting Gaussian quantum states still have a positive Wigner function.

Negativity of the Wigner function \cite{Kenfack:2004aa} has been identified as a necessary ingredient for implementing processes that cannot be simulated efficiently with classical resources \cite{mari_positive_2012,rahimi-keshari_sufficient_2016}, and is therefore an essential resource \cite{Takagi:2018aa,PhysRevA.98.052350} to achieve a quantum advantage. In networked quantum technologies it is, thus, crucial to generate and distribute Wigner-negativity in the nodes of a quantum network. In this spirit we focus here on the remote generation of Wigner-negativity, such that operations in one node of a quantum network create negativity in the Wigner function of another node while upholding the entanglement in the quantum network.

Photon subtraction \cite{dakna_generating_1997,Wenger:2004aa,parigi_probing_2007} is a natural candidate for such a non-destructive operation. In previous work, we showed that the subtraction of a photon causes an interplay between correlations and non-Gaussian features \cite{walschaers_entanglement_2017}. In the specific case of graph states, it was shown that photon subtraction propagates non-Gaussian properties through the system \cite{Walschaers:2018aa,Ra2019}, 
however it is far from clear whether this mediated non-Gaussianity has quantum features (see also \cite{Katamadze:2018gl}). Hence, here we turn the question around, and ask what type of correlations are required to remotely generate Wigner-negativity through photon subtraction. We show that Einstein-Podolsky-Rosen (EPR) steering \cite{Einstein:1935aa,Schrodinger:1935aa,Wiseman:2007aa,PhysRevA.80.032112,Reid:2009aa} is a necessary and sufficient condition.

In its essence, EPR steering focusses on the the structure of conditional quantum probabilities: when Alice and Bob share a quantum state, Bob can perform measurements on his part of the state and Alice can condition her state on Bob's measurement outcomes. Quantum correlations can induce an effect on Alice's statistics that cannot be explained by classical probability theory. If this is the case, Bob is said to steer Alice's state. In communication protocols, steering 
allows Alice to verify that strong quantum correlations are present and that Bob conveyed the correct measurement outcomes \cite{Branciard:2012aa,Kogias:2017aa,PhysRevLett.118.020402}. 
Motivated by such applications, EPR steering has been demonstrated in a range of experiments \cite{Ou:1992aa,Saunders:2010aa,Handchen:2012aa,Deng:2017aa,Cavailles:2018aa,Cai-Steering}. Our current results introduce the remote generation of Wigner-negativity as a new application of EPR steering, which is achievable with state-of-the-art experimental techniques.

Our manuscript is organised as follows: First we present a general derivation for the condition for remote preparation of Wigner negativity through photon subtraction, where we show that EPR steering is necessary. Then, we prove that, with an additional Gaussian transformation prior to photon subtraction, EPR steering becomes necessary and sufficient. Finally, we study a two-mode EPR state and a six-mode graph state as examples.\\

First of all, let us introduce the theoretical framework of multimode continuous-variable quantum optics \cite{Fabre2019}, by introducing the observable $\hat E(\vec r, t)$ that describes an $m$-mode electric field:
\begin{equation}
\hat E(\vec r, t) = \epsilon_c  \sum^m_{j=1}\frac{(\hat x_j + i \hat p_j)}{2}u_j(\vec r, t),
\end{equation}
where $\{u_1(\vec r, t), \dots, u_m(\vec r, t)\}$ is a mode basis, i.e.~an orthonormal set of solutions to Maxwell's equations that describe the system under consideration, and $\epsilon_c$ represents the single photon electric field. 
The observables $\hat x_j$ and $\hat p_j$ are known as the amplitude and phase quadratures, respectively, and follow the canonical commutation relations $[\hat x_j, \hat p_k] = 2i \delta_{j,k}$, $[\hat x_j, \hat x_k]=0$, and $[\hat p_j, \hat p_k]=0$.  

The mode basis $\{u_j(\vec r, t)\}$ is not unique, and can be changed through passive linear optics. When the mode basis is changed, the observables $\hat x_j$ and $\hat p_j$ change with it. Hence, it is often convenient to define the quadrature observable
$\hat q(f) = \sum_{j=1}^m (f_{2j-1} \hat x_j + f_{2j} \hat p_j),$
where $f \in \mathbb{R}^{2m}$  is a normalised vector in phase space, associated with a given mode. This leads to the general commutation relation $[\hat q(f_1), \hat q(f_2)] = -2i(f_1,\Omega f_2)$, where $(.,.)$ is the standard inner product and $\Omega$ is the symplectic form with properties $\Omega^2 = - \mathds{1}$, and $\Omega^t = - \Omega$. 

Our present work starts out from a multimode Gaussian state $\rho$, from which a photon is subtracted to render the state non-Gaussian. The resulting photon-subtracted state is given by
\begin{equation}
\rho^- = \frac{\hat a(g)\, \rho \hat a^{\dag}(g)}{\tr [\rho \, \hat a^{\dag}(g)\hat a(g)]},
\end{equation}
where $g \in \mathbb{R}^{2m}$ is associated with the mode in which we subtract the photon. The Wigner function of $\rho^-$ can be obtained analytically \cite{walschaers_entanglement_2017}, leading to a phase space representation of the state. Furthermore, the methods used to obtain this result, can equally be applied to find the Wigner function of a single mode, where all other modes are integrated out \cite{walschaers_statistical_2017}.

Here, we use correlations with the aim to remotely induce Wigner-negativity. Hence, we study the reduced Wigner function for a single mode $f \in \mathbb{R}^{2m}$, given that we subtracted the photon in an orthogonal mode $g \in \mathbb{R}^{2m}$. To express this single-mode Wigner function, we must introduce several objects. First, we define $V_f$ and $V_g$ as the two-dimensional covariance matrices \cite{RevModPhys.84.621} for modes $f$ and $g$, respectively. Moreover, we introduce the $2 \times 2$ matrix $V_{fg}$ that contains the correlations between these modes, such that
\begin{equation}\label{eq:Vjoint}
V_{\{f,g\}} = \begin{pmatrix} V_f & V_{fg}\\ V^t_{fg} & V_g\end{pmatrix},
\end{equation}
is the covariance matrix that describes the two-mode system associated with $f$ and $g$. Note that $V_{\{f,g\}}$ is a positive symplectic matrix with respect to the symplectic form $\Omega$ with $m=2$. Finally, we introduce $\xi_f, \xi_g \in \mathbb{R}^2$ that describe the mean field (i.e.~the displacement) in modes $f$ and $g$, respectively. After the subtraction of a photon, we find that the single-mode Wigner function for the mode $f$ is given by
\begin{align}\label{eq:WignerDispRed}
W_f^{-}(\beta_f)=  &\frac{\exp \left\{-\frac{1}{2} ([\beta_f - \xi_f], V_f^{-1} [\beta_f - \xi_f])\right\}}{2\pi \sqrt{\det V_f}(\tr\big[V_g \big] + \norm{\xi_g}^2 - 2)} \\&\times\Big\{ \norm{ V_{gf} V_f^{-1} (\beta_f - \xi_f) + \xi_g}^2 +\tr[V_{g\mid f}] - 2\Big\}\nonumber,
\end{align}
where $\beta_f \in \mathbb{R}^{2}$ denotes an arbitrary point in the two-dimensional phase space associated with mode $f$. 
In (\ref{eq:WignerDispRed}), we introduced the Schur complement of $V_{\{f,g\}}$, given by
 \begin{equation}V_{g\mid f} \equiv V_g - V_{fg}^t V_f^{-1}V_{fg}.\label{eq:condcovmat}\end{equation}
In statistical terms \cite{Muirhead:aa}, this matrix describes the variance of mode $g$ when conditioned to a specific joint measurement outcome for the phase- and amplitude quadrature in mode $f$ (a procedure which is unphysical). 

As a next step, we note that the Wigner function $W_f^{-}(\beta_f)$ is negative if and only if the polynomial $ \norm{ V_{gf} V_f^{-1} (\beta_f - \xi_f) + \xi_g}^2 +\tr[V_{g\mid f}] - 2$ is negative. This polynomial reaches its minimal value $\tr[V_{g\mid f}] - 2$ in the point $\beta_f = -V_f (V_{fg}^t)^{-1}\xi_g + \xi_f$. Hence, we find the simple negativity condition
\begin{equation}\label{eq:condition}
\tr[V_{g\mid f}]  < 2.
\end{equation}
Now we must understand the constraints that (\ref{eq:condition}) puts on the correlation in the initial Gaussian state. The relation between the Schur complement $V_{g\mid f}$ and EPR steering is crucial for this understanding. 

The mode $f$ is said to steer the mode $g$ when conditional covariances violate Heisenberg's uncertainty relation \cite{PhysRevA.40.913}, which is directly related to the properties of $V_{g\mid f}$ \cite{Wiseman:2007aa}.
More specifically, Gaussian EPR steering can be quantified through the Williamson decomposition which (in this single-mode case) allows to write $V_{g\mid f} = \nu S^tS$, where $S$ is a symplectic matrix and $\nu > 0$. In the absence of EPR steering, the positive matrix $V_{g\mid f}$ gives rise to values $\nu \geqslant1$. On the other hand, when mode $f$ can steer mode $g$, we find that $\nu < 1$. Note that $\nu$ can be used to define a measure for the capability of $f$ to steer $g$ \cite{Kogias:2015aa}. Using  this decomposition in (\ref{eq:condition}), we find that
\begin{equation}\label{eq:steerRes}
\nu \overset{\text{eq.}~(\ref{eq:condition})}{<} \frac{2}{\tr [S^t S]} \leqslant 1,
\end{equation}
where the latter inequality follows from the fact that $S^tS$ is a positive symplectic matrix (and thus has a trace larger than two).

Condition (\ref{eq:steerRes}) shows that {\em EPR steering is a necessary condition to remotely prepare Wigner-negativity through photon subtraction}. More specifically, if a photon is subtracted in mode $g$ and the single-mode Wigner function of a correlated mode $f$ is rendered negative, this implies that homodyne measurements in mode $f$ are able to steer homodyne measurements in mode $g$.\\

Although (\ref{eq:steerRes}) shows that EPR steering is a necessary condition, it is typically not a sufficient one. The condition entails that a certain amount of steering is required, and this amount depends on the fine details of the initial Gaussian state, as captured by the factor $\tr [S^t S]$. However, in the derivation of (\ref{eq:steerRes}) we only consider photon subtraction on mode $g$. It is thus natural to search for an additional operation in mode $g$ that can make EPR steering also a sufficient resource to remotely generate Wigner-negativity. This additional operation turns out to be a local Gaussian transformation. Note that applying such a local operation to the initial Gaussian state cannot change its steering properties.

We thus consider a local Gaussian operation on mode $g$, represented by a symplectic matrix $R$, which we implement prior to the subtraction of the photon. Because this operation is local, it only affects $V_g$ and $V_{fg}$, as given by $V_g \mapsto R^t V_g R$, and $V_{fg} \mapsto V_{fg}R$.
Hence, as this local operation is implemented prior to the photon subtraction, we can modify $V_g$ and $V_{fg}$ in (\ref{eq:WignerDispRed}) accordingly, and 
 (\ref{eq:condition}) then becomes $\tr[R^t V_{g\mid f} R]  < 2$.
When we apply Williamson's decomposition to this inequality, we find that (\ref{eq:steerRes}) changes as
\begin{equation}\label{eq:steerRes2}
\nu < \frac{2}{\tr [R^tS^t SR]}
\end{equation}
Thus, in contrast to EPR steering, the condition for remote creation of Wigner-negativity in mode $f$ is not independent of local Gaussian transformations on mode $g$, as shown directly by (\ref{eq:steerRes2}). Because we allow any local Gaussian operation $R$ on mode $g$, we can choose $R = S^{-1}$ and obtain our key result:
\begin{equation}
\text{Wigner-negativity in $f$} \overset{\text{eq.~(\ref{eq:steerRes2})}}{\iff} \nu < 1 \iff f \text{ steers } g.  \label{eq:steering}
\end{equation} 
{\em Thus, we can induce negativity in the Wigner function in mode $f$ by implementing local Gaussian operation and a photon subtraction in mode $g$ if and only if mode $f$ can steer mode $g$.} From (\ref{eq:steerRes2}), we see that the Gaussian transformation $R$ must change the squeezing for the effect to be non-trivial. Since the transformation is obtained through the Williamson decomposition of $V_{g\mid f}$, which is a priori not a physical covariance matrix, there is no simple relation between the required squeezing values in the local Gaussian transformation and the physical noise in mode $g$.\\

Remark that the conditions (\ref{eq:condition}) and (\ref{eq:steering}) are independent on the displacement of the state. Thus, in principle the displacement plays no role in whether or not $W_f^{-}(\beta_f)$ reaches negative values. However, our criteria only test the existence of Wigner-negativity, they do not provide a quantitative measure for this negativity. We can evaluate $W_f^{-}(\beta_f)$ in $\beta_f = -V_f (R^tV_{fg}^t)^{-1}\xi_g + \xi_f$ to find the minimal value $W_{\min}$ of the Wigner function
\begin{equation}\label{eq:wmin}
W_{\min} = \frac{(\tr[ R^tV_{g\mid f}R] - 2)e^{-\frac{1}{2} \left(\xi_g, [V_{fg}R]^{-1} V_f [R^tV_{fg}^t]^{-1}\xi_g\right)}}{2\pi \sqrt{\det V_f}\Big(\tr\big[R^tV_gR \big] + \norm{\xi_g}^2 - 2\Big)}.
\end{equation}
Here we see that $W_{\min}$ rapidly tends to zero for increasing displacements $\xi_g$. In other words, even though displacement does not play a role in (\ref{eq:condition}) and (\ref{eq:steering}) for determining whether the Wigner function is negative, the displacement $\xi_g$ in the mode $g$ of photon subtraction strongly impacts the amount of negativity that is induced in mode $f$. Surprisingly, the displacement $\xi_f$ in mode $f$ does not play any role in (\ref{eq:wmin}). Given that the Wigner-negativity in mode $f$ is manifested around $\beta_f = -V_f (R^tV_{fg}^t)^{-1}\xi_g + \xi_f$, this implies this scheme can be used to generate ``bright'' non-Gaussian states.

Another important element in determining the amount of Wigner-negativity in $W_f^{-}(\beta_f)$ is purity of mode $f$, given by $(\det V_f )^{-1/2}$ in (\ref{eq:wmin}). This highlights two competing effects: on the one hand, to induce Wigner-negativity in mode $f$, we require EPR steering between modes $g$ and $f$, which induces impurity in the reduced single-mode state of $f$. On the other hand, impurity of mode $f$ reduces the amount of Wigner-negativity that can be induced.\\

\begin{figure}
\centering
\includegraphics[width=0.49\textwidth]{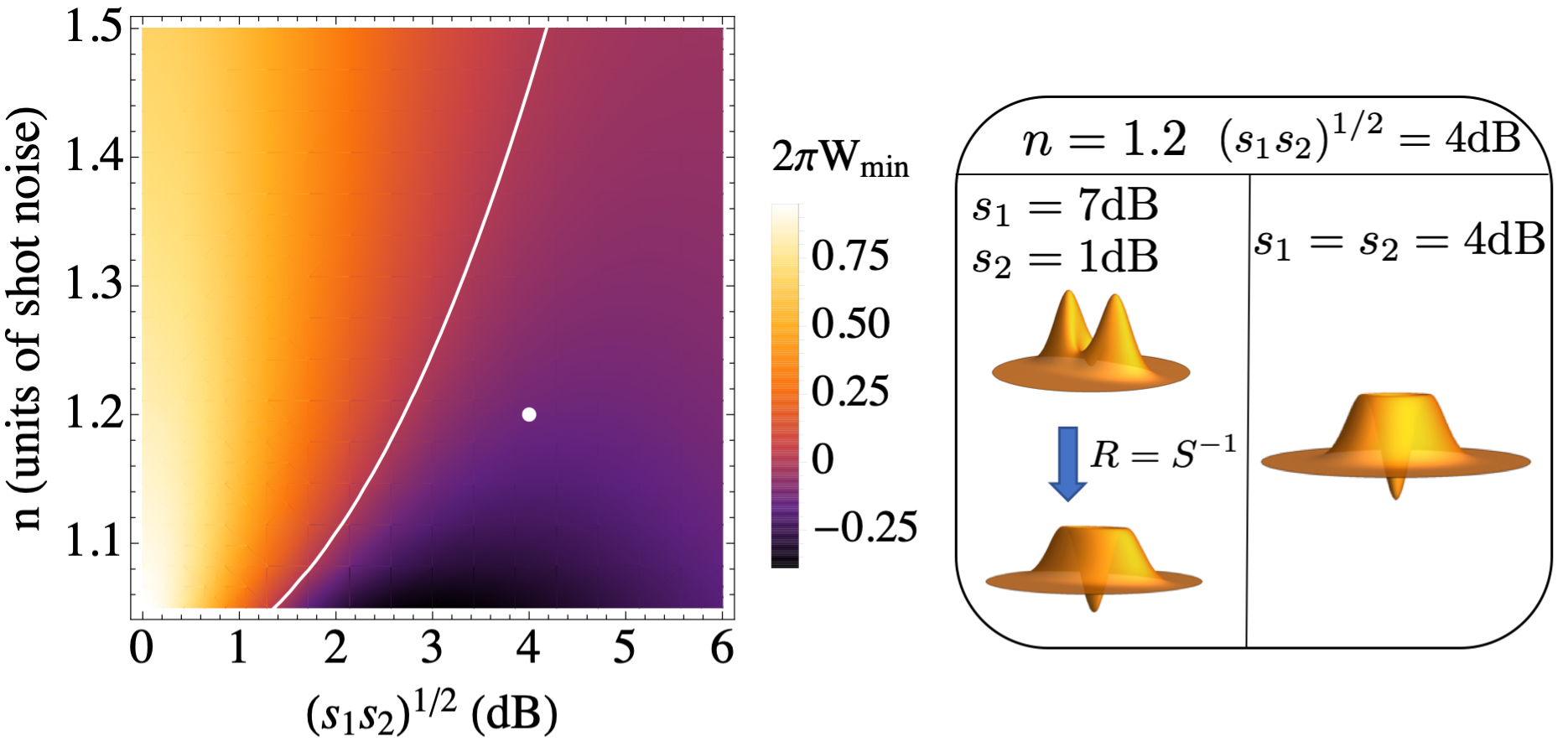}
\caption{{\bf Left Panel:} For the impure EPR state (see main text), remotely generated Wigner-negativity $W_{\min}$ (\ref{eq:wmin}), with $R=S^{-1}$, is shown as a function of thermal noise $n$ and the geometric mean of the squeezing parameters $(s_1 s_2)^{1/2}$. The white curve indicates $\nu = 1$. The white dot indicates the parameters $n=1.2$ and $(s_1 s_2)^{1/2} = 4 {\rm dB}$, with $2\pi W_{\rm min} \approx -0.135$, which is studied explicitly in the {\bf Right Panel}: Different setup with $(s_1 s_2)^{1/2} = 4 {\rm dB}$ are shown. When $s_1 = 7 {\rm dB}$ and $s_2= 1 {\rm dB}$ condition (\ref{eq:condition}) is not fulfilled, but a local Gaussian transformation $R=S^{-1}$ allows to reach $2\pi W_{\rm min} \approx -0.135$. For $s_1 = s_2= 4 {\rm dB}$, condition (\ref{eq:condition}) is satisfied, and $2\pi W_{\rm min} \approx -0.135$ is achieved without local Gaussian transformation. Note that the Wigner function has the property that $-1 < 2\pi W_{\rm min} < 1$. \label{fig:Example1}}
\end{figure} 

Up to this point, the argumentation held for any arbitrary pair modes $f$ and $g$, let us now make the obtained results more concrete. First of all, we consider the EPR state  \cite{Einstein:1935aa,PhysRevA.40.913,PhysRevLett.60.2731,Ou:1992aa,PhysRevA.58.4345}, which is obtained by mixing two squeezed vacuum states with covariance matrices
\begin{equation}
V_1 = \begin{pmatrix} n s_1 & 0\\ 0 &n/s_1\end{pmatrix}, \quad \text{and}\quad V_2 = \begin{pmatrix} n /s_2 & 0\\ 0 &ns_2\end{pmatrix}
\end{equation}
on a balanced beamsplitter. Here we assume the presence of some additional thermal noise $n$, which is the same in each mode. We create a negative Wigner function in one of the two EPR-modes (mode $f$) by subtracting a photon in the other (mode $g$). 

To assess EPR steering, $V_{g\mid f}$ can be calculated in a straightforward way from which we then obtain
\begin{equation}\label{eq:nuEPR}
\nu = \frac{2 n (s_1 s_2)^{1/2}}{1+s_1s_2}.
\end{equation}
Note that $\nu$, and therefore also the capacity to remotely prepare Wigner-negativity according to (\ref{eq:steering}), does not depend on the individual values of $s_1$ and $s_2$, but only on their geometric mean  $(s_1 s_2)^{1/2}$. Computing the minimal value of the Wigner function (\ref{eq:wmin}), we observe that it shares this dependence on $(s_1 s_2)^{1/2}$, under the condition that $R = S^{-1}$. This minimal value is plotted in the left panel of Fig.~\ref{fig:Example1}, and reaches significantly negative values (comparable to what is experimentally achieved in \cite{Ra2019} for instance) for realistic amounts of thermal noise and squeezing. The white curve corresponds to the case where $\nu = 1$ and $W_{\rm min} = 0$, and can be associated with the experimental scenario of having $50\%$ losses in the system. Furthermore, we observe that $W_{\rm min}$ tends towards zero as the squeezing parameter is increased, which is a consequence of an associated increase in entanglement. Larger entanglement leads to a lower purity in modes $f$ and $g$, thus increasing the denominator in (\ref{eq:wmin}).

The right panel of Fig.~\ref{fig:Example1} explicitly shows the example for $n=1.2$ and $(s_1 s_2)^{1/2} = 4 {\rm dB}$. In the symmetric case with $s_1 = s_2 = 4{\rm dB}$ (experimentally realised with $\sim 5 {\rm dB}$ squeezing, subjected to $\sim 30\%$ losses) condition (\ref{eq:condition}) is satisfied and $S = \mathds{1}$, such that nothing can be gained by performing a Gaussian operation prior to photon subtraction.  We therefore consider this to be the optimal case. However, $(s_1 s_2)^{1/2} = 4 {\rm dB}$ can also be achieved by very different choices of parameters, e.g., $s_1 = 7 {\rm dB}$ and $s_2 = 1{\rm dB}$. In this case we explicitly show that no Wigner-negativity can be generated remotely with only photon subtraction, i.e., condition (\ref{eq:condition}) is not satisfied. However, by implementing a local Gaussian transformation $R = S^{-1}$ in mode $g$, we can fulfil (\ref{eq:steering}) and reach a significant amount of Wigner-negativity equal to the optimal case $ W_{\rm min} \approx -0.135/2\pi$.
This example shows that the main role of the local Gaussian transformation $R$ is to balance the noise in modes $f$ and $g$. This explains why the symmetric setup with $s_1 = s_2 = 4{\rm dB}$ is the optimal case. \\

Impure two-modes states do not only arise due to losses, they could also originate from entanglement to additional modes. To explicitly explore this case, we now subtract a photon from a mode in a larger multimode state. In particular, we consider CV graph states \cite{Su:12,PhysRevLett.112.120505,cai-2017,Larsen:2019aa,Asavanant:2019aa}, which form the backbone of measurement-based quantum computing in CV \cite{gu_quantum_2009}, and have tractable entanglement properties. Recently, EPR steering was experimentally observed in such a system \cite{Deng:2017aa}. These states are Gaussian, with a covariance matrix that is built in accordance with a graph ${\cal G}$ as a blueprint. The graph ${\cal G}$ describes the entanglement pattern that will be imprinted on a set of squeezed modes. For simplicity, we assume that all modes are equally squeezed. Now, let ${\cal A}$ denote the adjacency matrix of the graph (i.e.~${\cal A}_{jk} = 1$ when $j$ and $k$ are connected, and zero otherwise), which gives us the final graph state covariance matrix \cite{PhysRevA.76.032321,Ferrini:2015aa}
\begin{equation}\label{eq:GraphState}
V = \begin{pmatrix}X & Y \\ -Y & X\end{pmatrix} \left[ \bigoplus_{j=1}^m  \begin{pmatrix}ns & 0 \\ 0 & n/s\end{pmatrix} \right] \begin{pmatrix}X & -Y \\ Y & X\end{pmatrix}, 
\end{equation}
where $X = ({\cal A}^2 + \mathds{1})^{-1/2}$ and $Y = {\cal A}X$. We refer to $s >0$ as the squeezing parameter, and $n>1$ denotes the fraction of added thermal noise.

\begin{figure}
\centering
\includegraphics[width=0.48\textwidth]{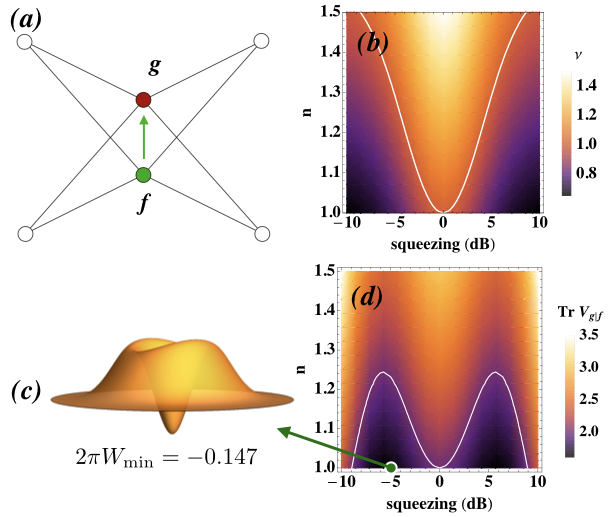}
\caption{Photon subtraction in mode $g$ of six-mode graph state (a) can render the Wigner function in mode $f$ negative [see Wigner function in Panel (c)]. For a graph state, generated according to (\ref{eq:GraphState}), the squeezing $s$ and the fraction of thermal noise $n$ (compared to vacuum noise) is varied, and we show how this influences the validity of condition (\ref{eq:steering}) in Panel (b), and condition (\ref{eq:condition}) in Panel (d). The white curves in panels (b) and (d) represent the case where $\nu = 1$ and $\tr[V_{g\mid f}]  = 2$, respectively. The arrow between $f$ and $g$ indicates direction of EPR steering.\label{fig:Example2}}
\end{figure} 

In Fig.~\ref{fig:Example2}, we consider a six-mode scenario, corresponding to a graph ${\cal G}$, as shown in panel (a). 
In panel (d), we see that (\ref{eq:condition}) is fulfilled in regions of intermediate squeezing, whereas the amount of thermal noise that can be allowed depends on the squeezing. EPR steering can be achieved for a larger range of parameters, as shown in panel (b), and thus there is something to be gained from local Gaussian transformations $R$. A priori, this is surprising, since the considered setup is highly symmetric (recall that in Fig.~\ref{fig:Example1} the local Gaussian transformation served to change the asymmetry between modes $f$ and $g$). This highlights a second important feature $R$: not only can it change the local squeezing, it can also change its orientation in phase space. The latter feature is crucial for the remote preparation of Wigner-negativity in this example. 

Panel (c) of Fig.~\ref{fig:Example2} explicitly shows that, even in systems with a significant number of modes, Wigner-negativity can be remotely prepared. This result is intriguing since additional entangled modes lead to more noise in the subsystem of modes $f$ and $g$. Naturally, for increasing values of thermal noise $n$, the Wigner-negativity in mode $f$ will steadily decrease. 

In conclusion, we have shown that photon subtraction can be used to transfer Wigner-negativity to an entangled node, given that condition (\ref{eq:condition}) applies: for photon subtraction in mode $g$ to create Wigner-negativity in mode $f$, it is necessary that $g$ is EPR steerable by $f$. EPR steering also becomes a sufficient condition when we allow for additional local Gaussian transformations prior to photon subtraction. In Figs.~\ref{fig:Example1} and \ref{fig:Example2}, we show that this remote generation of negativity is feasible with realistic values of squeezing and thermal noise, and in relevant multimode states such as graph states. This result highlights the potential use of EPR steering as a tool for studying and Wigner-negativity and vice-versa. For instance, in multimode photon subtraction experiments, it is a priori hard to witness Wigner negativity due to the complexity of multimode tomography. For some states, our result provides an elegant way of identifying exactly which individual mode needs to be measured as a witness for this Wigner-negativity, and in other cases our result provides a way to prove that any present Wigner-negativity cannot be witness only through single-mode measurements.


\begin{acknowledgements}
{\em Acknowledgements ---} We thank Qiongyi He, Yu Xiang, and Fr\'ed\'eric Grosshans for very inspiring and fruitful discussions. M.W. was partially funded through research fellowship WA 3969/2-1 from the German Research Foundation (DFG). N.T. acknowledges financial support of the Institut Universitaire de France. 
\end{acknowledgements}
\bibliography{notes_steering.bib}

\end{document}